# The contribution of pattern recognition of seismic and morphostructural data to seismic hazard assessment

A. Peresan[1,3,5], A. Gorshkov[1,2], A. Soloviev[1,2], G.F. Panza[1,3,4,5]

[1] *The Abdus Salam International Centre for Theoretical Physics, SAND Group, Trieste, Italy*
[2] *IEPT of Russian Academy of Sciences, FASO, Moscow, Russian Federation*
[3] *Department of Mathematics and Geosciences, University of Trieste, Trieste, Italy.*
[4] *Institute of Geophysics, China Earthquake Administration, Beijing*
[5] *International Seismic Safety Organization (ISSO)*
*e-mail: aperesan@units.it*

## Abstract

Experience from the destructive earthquakes worldwide, which occurred over the last decade, motivated an active debate discussing the practical and theoretical limits of the seismic hazard maps based on a classical approach (PSHA). Systematic comparison of the observed ground shaking with the expected one, in fact, shows that such events keep occurring where PSHA predicted seismic hazard to be low. Amongst the most debated issues is the reliable statistical characterization of the spatial and temporal properties of large earthquakes occurrence, due to the unavoidably limited observations from past events.

We show that pattern recognition techniques allow addressing these issues in a formal and testable way and thus, when combined with physically sound methods for ground shaking computation, like the neo-deterministic approach (NDSHA), may produce effectively preventive seismic hazard maps.

Pattern recognition analysis of morphostructural data provide quantitative and systematic criteria for identifying the areas prone to the largest events, taking into account a wide set of possible geophysical and geological data, whilst the formal identification of precursory seismicity patterns (by means of CN and M8S algorithms), duly validated by prospective testing, provides useful constraints about impending strong earthquakes at the intermediate space-time scale. According to a multi-scale approach, the information about the areas where a strong earthquake is likely to occur can be effectively integrated with different observations (e.g. geodetic and satellite data), including regional scale modeling of the stress field variations and of the seismic ground shaking, so as to identify a set of priority areas for detailed investigations of short-term precursors at local scale and for microzonation studies.

Results from the pattern recognition of earthquake prone areas ($M \geq 5.0$) in the Po plain (Northern Italy), as well as from prospective testing and validation of the time-dependent NDSHA scenarios are presented, including the case of the May 20, 2012 Emilia earthquake.

## Introduction

Over the last decade, the large and most destructive earthquakes worldwide kept occurring in areas where the seismic hazard predicted by classical PSHA maps (Giardini et al., 1999) was comparatively lower (Wyss et al., 2012; Stein et al. 2012, Geller et al., 2011). The unsatisfactory anticipatory capability of PSHA maps motivated an active debate discussing their practical and theoretical limits (e.g. Peresan et al. 2012; Gulkan, 2013; Stein et al., 2013 and references therein).

Amongst the raised questions, several point to the limits in the reliable characterization of the spatial and temporal properties of potential earthquake sources, namely:

- Is the available information about past seismicity sufficient to constrain the areas where large earthquakes may occur?
- Does the available information about past seismicity allow us to constrain the statistical models and to provide reliable estimates of their probability of occurrence within narrow regions?

Recent destructive earthquakes suggest this might not be the case; most of them, indeed, occurred in areas where similar events were not reported before. This observation is particularly critical when dealing with the classical probabilistic (PSHA) approach to seismic hazard assessment, which requires the definition of earthquake probability models within narrow seismic sources, often resorting to simplistic and untestable assumptions, poorly constrained by available observations (e.g. poissonian earthquakes occurrence, characteristic earthquake model, etc).

The heuristic limitations are, indeed, a major limit of PSHA - the available short earthquake catalogues worldwide do not allow the statistics inference theory to project the present time probabilistic estimation over periods of time as long as 1000-10000 years. Clearly, the map tries to predict events, which do not have a significant statistics within the input data, and most probably many of them did not even appear yet, accordingly with the characteristic time of geological process leading to seismogenesis. This calls for the integration of the information provided by earthquake catalogues with additional geological and geophysical evidences, independent from observed seismicity. However, with PSHA it is quite difficult, if not impossible, to take in due account the information given by paleoseismicity (e.g. Michetti et al., 2012), morphostructural analysis (Gorshkov et al., 2003) or similar studies (e.g. Brune and Whitney, 2000). In fact, though geological and geophysical evidences may contribute to identify areas where large, yet unobserved earthquakes may occur, defining the associated probability distribution is by far a much more difficult problem, particularly in view of the large uncertainties and incompleteness characterizing such information.

We show in this paper that pattern recognition techniques allows us addressing the above questions in a formal and testable way and thus, when combined with physically sound methods for ground shaking computation, like the neo-deterministic approach (NDSHA, Panza et al., 2012), may significantly contribute to the production of effectively preventive seismic hazard maps.

The first application of pattern recognition to the identification of potential locations of strong earthquakes has been introduced in 1970s by Gelfand et al. (1972). The approach is based on the assumption that large earthquakes are correlated with the nodes formed around lineaments intersections. The nodes are delineated by morphostructural zoning method (MZ) that is based on geomorphic and geological data and does not rely on the knowledge about past seismicity (Alexeevskaya et al., 1977). The pattern recognition technique is employed to select among all the nodes those, which are prone to earthquakes above a certain (target) magnitude. Since the early 1970s the methodology has been successfully applied in a number of regions worldwide, including California, where it permitted the identification of earthquake prone areas that have been subsequently struck by strong events and that previously were not considered seismogenic (see the overviews by Gorshkov et al., 2003; Soloviev et al., 2014). The reliability and statistical significance of the methodology is supported by the number of strong earthquakes that occurred in the previously studied regions: 87% of the post-publication events with relevant magnitudes occurred within the nodes that were recognized in advance as capable of target earthquakes (Soloviev et al., 2014).

The morphostructural analysis has been already applied to the Italian mountain ranges, namely in the Alps (Gorshkov et al., 2004) and in peninsular Italy and Sicily (Gorshkov et al., 2002). In this study the analysis is applied for the identification of earthquake prone areas ($M \geq 5.0$) in the Po plain (Northern Italy), following the specific formal criteria of MZ recently defined for plain areas, where special attention is paid to drainage pattern (variant $MZ_P$). The feasibility of such analysis ($MZ_P$) has been established for topographically flat basins within Iberian Peninsula (Gorshkov et al., 2010) and recently for the Rhone Valley in France (Gorshkov and Gaudemer, 2012). In 2012 the southeastern part of the Po plain, Emilia-Romagna region, was struck by a $M$=6.1 earthquake, which occurred within one of the nodes delineated by $MZ_P$; we show that, if the recognition would have been performed any time before the event after the formalization of $MZ_P$ the node hosting the 20 May 2012 Emilia Romagna earthquake could have been properly recognized as prone to $M \geq 5.0$ earthquakes.

A formal identification of priority areas, most prone to future seismic activation, can be performed integrating the space information provided by the earthquake prone areas with the space-time information provided by formally defined earthquakes predictions (e.g. Peresan et al., 2011). A number of forecasting methodologies have been developed, mainly based on the analysis of seismicity, and are currently applied both at national and international scale (Jordan et al., 2011). However, only few of them, including CN and M8S algorithms (Peresan et al., 2005), are defined in testable way and are undergoing rigorous prospective testing over a time span sufficient to validate, in a statistically reliable way, the obtained results. The algorithms make use of the information provided by standard earthquake catalogs to identify the periods of time when a strong earthquake (with magnitude above a predefined threshold) is likely to occur within a region with linear dimensions of few hundred km. CN and M8S, which are based on the diagnosis of formally defined premonitory seismicity patterns, have demonstrated effective and statistically significant results in rigorous real-time testing, ongoing for more than a decade over the Italian territory (Peresan et al., 2011) and about two decades on a global scale (Kossobokov, 2012).

The indications provided by pattern recognition procedures can be used to focus the investigation of possible local scale precursors in the areas (with linear dimensions of few tens kilometers), where the probability of a strong earthquake is relatively high, as well as for drawing time-dependent scenarios of ground shaking by the NDSHA approach (Peresan et al., 2011). Accordingly, a set of deterministic scenarios of ground motion at bedrock, which refers to the time interval when a strong event is likely to occur within the alerted area, can be defined by means of full waveform modeling, both at regional and local scale.

In Italy CN and M8S predictions, as well as the related time-dependent ground motion scenarios associated with the alarmed areas, are routinely updated every two months since 2006. The prospective application of the time-dependent NDSHA approach provides information that can be useful in assigning priorities for timely mitigation actions and, at the same time, allows for a rigorous prospective testing and validation of the proposed methodology. As an example, in the areas where ground shaking values greater than 0.2 g are estimated at bedrock, further investigations can be performed taking into account the local soil conditions, to assess the performances of relevant structures, such as historical and strategic buildings.

## 1. State of art in pattern recognition of earthquake prone areas

A systematic and testable assessment, capable of providing first-order consistent information about the sites where large earthquakes may occur, is highly important for knowledgeable seismic hazard evaluation. One of the approaches to the solution of this problem was introduced in the early 1970s by Gelfand et al. (1972). The approach is based on the hypothesis that the strong earthquakes nucleate at the morphostructural nodes, namely the intersections of morphostructural lineaments, the nodes. Here the term "lineament", as will be described in section 1.1, is used with a quite different meaning with respect the early one introduced by Hobbs (1904). Within the study region an earthquake is defined as "strong" if its magnitude $M \geq M_0$, where $M_0$ is a threshold specified depending on the seismicity level of the region. The nodes are delineated by the specific morphostructural zoning (MZ) method (Alekseevskaya et al., 1977; Rantsman, 1979; Gorshkov et al., 2003). The hypothesis that the epicenters of the strong earthquakes correlate with the lineaments intersections is supported by the statistical analysis of their locations, performed by Gvishiani and Soloviev (1981). The few exceptions to this hypothesis can be explained by the possible errors in both determining the location and size of the earthquakes and in the mapping of the morphostructural nodes.

Since the systematic instrumental seismological observations dates back only about 100 years and the available earthquake catalogs cover a time span rather short compared to characteristic time of tectonic processes, it is reasonable to assume that not all the potentially seismogenic nodes are marked by strong earthquakes during the period covered by instrumental and macroseismic observations. For a partial but effective solution of this problem, Gelfand et al. (1972) proposed to apply pattern recognition methods, specifically the Cora-3 algorithm with learning designed by Bongard (1967). The nodes are treated as the objects of recognition; each object is associated with the vector of topographic, geological and geophysical parameters describing nodes. The sample nodes needed for training the pattern recognition algorithm are selected taking into account the data about seismicity recorded in the region. The output of the recognition includes: (1) the classification of the nodes into those prone to earthquakes with $M \geq M_0$ and those, where only the events with $M < M_0$ are possible, and (2) the criteria (the decision rule) governing the classification. The criteria are based on topographic, geological, and geophysical parameters characterizing the nodes.

## 1.1. Pattern recognition applied to earthquake prone areas identification

The recognition of earthquake prone areas includes the following steps:
• outline of the region and specification of the magnitude $M_0$ which defines the "strong" earthquake for a given region;
• morphostructural zonation of the region and specification of the objects of recognition;
• selection of the sample sets used for the training of the pattern recognition algorithm;
• selection of the parameters for the description of the objects of recognition and measurement of their values;
• discretization of the values of the parameters and their convertion into binary vectors;
• application of a pattern recognition algorithm in order to classify the objects of recognition into two classes: the seismogenic objects prone to $M \geq M_0$ and non-seismogenic ones with respect to $M_0$;
• estimation of the reliability of the classification by control tests.

When selecting the region for the study, two main factors are taken into account. Firstly, the region should be relatively homogeneous with respect to the tectonic regime governing the seismogenesis in the region. Secondly, the number of recorded events with target magnitude should be sufficient to form the representative sample set for the training of the pattern recognition algorithms. Once the region is specified, the MZ is carried out to delineate the morphostructural nodes, which are then treated as recognition objects.

Since 1972 the methodology has been applied for the recognition of earthquake prone areas for different $M_0$ in many seismic regions of the world, specifically, in the Tien Shan and Pamir (Gelfand et al., 1972), united region of Balkans, Asia Minor, Transcaucasia (Gelfand, 1974a; 1974b), California and Nevada (Gelfand et al., 1976), Italy (Caputo et al., 1980; Gorshkov et al., 2002), Andean South America (Gvishiani and Soloviev, 1984a), Kamchatka (Gvishiani et al., 1984b), Western Alps (Cisternas et al., 1985), Pyrenees (Gvishiani et al., 1987), Greater Caucasus (Gvishiani et al., 1988), Lesser Caucasus (Gorshkov et al., 1991), Himalaya (Bhatia et al., 1992), Carpathians (Gorshkov et al., 2000), Alps and Dinarides (Gorshkov et al., 2004), Alborz (Gorshkov et al., 2009a), Ecuador (Chunga et al., 2010), Iberian Plate (Gorshkov et al., 2010), North Vietnam (Tuyen et al., 2012), Kopet Dagh (Novikova and Gorshkov, 2013), Caucasus Region (Soloviev et al., 2013). The other applications of pattern recognition techniques for seismic hazard analysis were recently reviewed by Mridula et al. (2013).

***Morphostructural zoning method.*** In the MZ (Alekseevskaya et al., 1977; Gorshkov et al., 2003) the study region is divided into a system of hierarchically ordered areas, characterized by homogeneous present-day topography and tectonic structure. MZ distinguishes (1) blocks (areas) of different rank; (2) their boundary zones, morphostructural lineaments; and (3) sites where lineaments intersect, the nodes. A morphostructural lineament is viewed as a boundary zone between territorial units delineated by MZ. The rank of the lineament depends on the rank of the area limited by the lineament. With respect to the regional trend of the tectonic structure and topography, two types of lineaments are distinguished: (1) longitudinal and (2) transverse ones. Longitudinal lineaments are approximately parallel to the regional strike of the tectonic structure and of the topography and, as a rule, include the prominent faults. Transverse lineaments go across the regional trend of the tectonic structure and of the topography. Normally, they appear on the Earth's surface discontinuously and are evidenced by escarpments, by rectilinear parts of river valleys, and partly by faults.

The MZ method differs from the standard morphostructural analysis where the term "lineament" (Hobbs, 1904) is used to define the complex of alignments detectable on topographic maps or on satellite images. According to that definition the lineament is locally defined and the existence of the lineament does not depend on the surrounding areas. In MZ, the primary element is the block – a relatively homogeneous area - while the lineament is a secondary element of the morphostructure. The boundaries of the blocks correspond to the lineaments; this means that the existence and the position of the lineaments are determined not locally, but as a part of a broader hierarchical structure. If a certain alignment does not separate two topographically different areas, that alignment cannot be viewed as a lineament in MZ; therefore, the lineaments are secondary features with respect to the blocks.

Since MZ is based mainly on topographic data, the method is applicable for studying practically any areas of the Earth including sea and oceanic floors if topographic and/or bathymetric data are available for the study region. The only exceptions are the desert areas where mobile aeolian landforms prevail. The first attempt to study the sea floor with MZ has been made by Novikova et al. (2012) for the deep South Caspian basin. The MZ map of the basin has been used for recognizing seismogenic nodes prone to M6+.

Morphostructural nodes are formed around the intersections or junctions of two or more lineaments. A node may include more than one intersection or junction. Nodes are characterized by a mosaic combination of various topographic forms and by an increased number of linear topographic forms of various strikes that reveal the instability of the area.

The methodology treats the nodes as earthquake-controlling structures. Nodes are the areal structures, boundaries of which can be delineated through long-term field work, as it has been proven in the Central Asia and the Greater Caucasus (Rantsman, 1979; Gvishiani et al., 1988). However, the delineation of the boundaries of the nodes is a cumbersome task and it requires large-scale MZ of lineament intersection areas based on field studies. An attempt to delineate structurally bounded nodes has been made by Gorshkov et al. (2009b) in the Alps-Dinarides junction zone on the base of the analysis of large-scale topographic and geological maps. But in most previously studied regions the intersections of morphostructural lineaments were treated as recognition patterns. In such case the nodes are defined as the circles of a certain radius, which depends on the target $M_0$ in accordance with the empirical relation between magnitude and the earthquake source (Wells and Coppersmith, 1994).

The fact that earthquakes are nucleated within the nodes was first established from the field geomorphic observations in Central Asia by Gelfand et al. (1972) and then it was proved in all other regions studied with this methodology (Gelfand et al., 1974a, 1974b, 1976; Caputo et al., 1980; Gorshkov et al., 1991, 2000, 2002, 2004, 2009a, 2010; Cisternas et al., 1985; Gvishiani et al., 1984a, 1984b, 1987, 1988; Bhatia et al., 1992; Chunga et al., 2010; Tuyen et al., 2012; Novikova et al., 2013; Soloviev et al., 2013).

The correlation of earthquakes with fault zones intersections in different tectonic environments was also evidenced by Talwani (1988), Hudnut et al. (1989), Girdler and McConnell (1994). The model proposed by Talwani (1999) demonstrates that intersecting fault zones provide a location for stress accumulation. According to King (1986), fault zones intersections provide locations for the initiation and healing of ruptures. A model proposed by Gabrielov et al. (1996) implies that block interaction along intersecting lineaments leads to stress and strain accumulation and secondary faulting around the intersection. This causes the generation of new faults and blocks of progressively smaller size, so that a hierarchical mosaic structure, essentially a node, is formed around the intersection.

***Recognition of earthquake prone areas.*** Since strong earthquakes are spatially correlated with nodes possessing some specific features, the identification of earthquake prone areas can be formulated as a pattern recognition problem, where the nodes are the objects of recognition. Specifically, the pattern recognition algorithm is applied in order to classify the nodes into two classes: **D** ("dangerous") nodes, which can host earthquakes with $M \geq M_0$, and the nodes **N** ("not dangerous") where earthquakes with $M \geq M_0$ cannot occur.

The application of recognition algorithms with learning requires the preliminary selection of a training sample set $\mathbf{W_0}$. This set consists of two non-overlapping subsets of objects with known classification, namely: the $\mathbf{D_0}$ objects that a priori belong to class **D**, and the $\mathbf{N_0}$ objects that a priori belong to class **N**. In this specific application the sample $\mathbf{W_0} = \mathbf{D_0} \cup \mathbf{N_0}$ is constructed in the following way. The subset $\mathbf{D_0}$ is composed of nodes that are already marked by earthquakes with $M \geq M_0$. The subset $\mathbf{N_0}$ is either composed by the remaining nodes of **W**, $\mathbf{N_0} = \mathbf{W} \backslash \mathbf{D_0}$, or by the nodes were smaller events, with $M \geq M_0 - \delta$ ($\delta$ is typically about 0.5), are unknown. We note that it is practically impossible to select an "uncontaminated" training set $\mathbf{N_0}$ for class **N**, since it is impossible to guarantee a priori that a node assigned to $\mathbf{N_0}$ is certainly an aseismic one.

Each node is associated with a vector. The recognition algorithms are applied to the vectors, whose components are given by the values of the parameters describing the nodes. Parameters of the nodes should characterize some properties of the node environments that reflect the different factors relevant to seismicity. During the studies on recognition of earthquake prone areas prone (Gelfand et al., 1972, 1974a, 1974b, 1976; Gorshkov et al., 1991; 2000, 2002, 2003, 2004, 2009a, 2010; Caputo et al., 1980; Gvishiani and Soloviev, 1984a; Cisternas et al., 1985; Gvishiani et al., 1988; Bhatia et al., 1992) different parameters describing the nodes have been tested. These parameters include the geomorphic and morphometric information, the characteristics of the block-and-lineament geometry, and the gravity data. In principle, any information characterizing the specific features of seismic areas can be used for recognition. The only necessary precondition is that the value of each parameter should be defined with the same accuracy for all the objects (i.e. the nodes) within the studied territory.

Once the values of the parameters have been specified, all the objects contained in **W** are converted into the vectors $\mathbf{w}^i = \{w_1{}^i, w_2{}^i, ..., w_m{}^i,\}$, $i = 1, 2, ..., n$, where $m$ is the total number of the parameters; $n$ is the total number of the objects in the set **W**; $w_k{}^i$ is the value of the $k$th parameters measured for the $i$th object. Normally, the Cora 3 recognition algorithm (Bongard, 1967) is used for the classification of nodes. The algorithm operates with binary vectors. Therefore, the initial values of the parameters should be converted into binary vectors. The conversion is carried out by the procedures of discretization and coding (Gelfand et al., 1976; Gvishiani et al., 1988; Gorshkov et al., 2003). After the conversion, the recognition algorithm carries out the classification of the objects using the training sample $\mathbf{W_0} = (\mathbf{D_0}, \mathbf{N_0})$: $\mathbf{W} = \mathbf{D} \cup \mathbf{N}$, where **D** and **N** are the vectors assigned by the algorithm to the classes **D** and **N**, respectively. In every detail the Cora 3 algorithms and some others used for recognition of earthquake prone areas are described in (Gelfand et al., 1976; Gorshkov et al., 2003).

The quality of the final classification defined by the recognition algorithm can be evaluated by the control tests. A number of such tests have been developed to estimate the stability of the final classification (Gelfand et al., 1976; Gvishiani et al., 1988; Gorshkov et al., 2003). The positive results of the control tests suggest that the objects of recognition are properly subdivided into the **D** and **N** classes. Evidences about historical and paleo-earthquakes are also taken into account in the evaluation of the final classification of the nodes. Clearly, the final response about the reliability of the recognition results obtained for the studied region may only come by prospective testing, accounting for the subsequent target earthquakes occurred after the nodes classification.

## 1.2 Validation of the results of nodes recognition

During last 40 years the recognition of earthquake prone areas has been performed for a number of seismically active regions (Gelfand et al., 1972, 1974a, 1974b, 1976; Caputo et al., 1980; Gorshkov et al., 1991, 2000, 2002, 2004, 2009a, 2010; Cisternas et al., 1985; Gvishiani et al., 1984a, 1984b, 1987, 1988; Bhatia et al., 1992; Chunga et al., 2010; Tuyen et al., 2012; Novikova and Gorshkov, 2013; Soloviev et al., 2013). Most of the regions have been struck by the post-publication earthquakes of target magnitudes. These events provide the necessary information to assess the reliability of the results. The verification of the recognition results using the post-publication earthquakes has been periodically performed (Gorshkov et al., 2003; 2005). The last verification has been made by Soloviev et al. (2014) using the earthquake parametric data provided by the U.S. National Earthquake Information Center (NEIC) as of August 1, 2012. Table 1 summarizes the results updated to April 2014. The table includes only the regions which were struck by the post-publication target events, with magnitudes $M \geq M_0$. In Italy, where the nodes recognition has been revised and extended during the last decade (Gorshkov et al., 2002, 2004), we refer to the early publication by Caputo et al. (1980). In fact the results, for earthquakes that occurred in peninsular Italy and Sicily after the publication of the papers, are the same in both Gorshkov et al. (2002) and Caputo et al. (1980), evidencing the stability of recognition, whereas no target earthquake did occur since 2004 in the Alpine region analyzed by Gorshkov et al. (2004).

Noticeably, Table 1 includes the most recent $M$=8.2 earthquake (according to NEIC determinations), which occurred in Chile on April 2, 2014; the epicentre of this earthquake is located within a node that was recognized as prone to $M \geq 7.75$ by Gvishiani and Soloviev (1984a).

**Table 1. Summary results of testing the reliability of recognition of the earthquake prone areas**

| Region and year of the publication of the results | $M_0$ | Number of post-publication target earthquakes in the region | | | |
|---|---|---|---|---|---|
| | | total number | in the **D** nodes (including in the **D\*** nodes) | in the **N** nodes | outside the nodes |
| Tien Shan and Pamir, 1972 | 6.5 | 7 | 6 (1) | 0 | 1 |
| Balkans, Asia Minor, and Transcaucasia, 1974 | 6.5 | 27 | 25 (7) | 1 | 1 |
| California and Nevada, 1976 | 6.5 | 14 | 13 (4) | 0 | 1 |
| Italy, 1980 | 6.0 | 8 | 4 (1) | 0 | 4 |
| South American Andes, 1982 | 7.75 | 6 | 5 (3) | 1 | 0 |
| Kamchatka, 1984 | 7.75 | 1 | 1 | 0 | 0 |
| Western Alps, 1985 | 5.0 | 6 | 5 (1) | 1 | 0 |
| Pyrenees, 1987 | 5.0 | 6 | 5 (1) | 1 | 0 |
| Greater Caucasus, 1988 | 5.0 | 13 | 13 (9) | 0 | 0 |
| Himalaya, 1992 | 6.5 | 4 | 3 (1) | 0 | 1 |
| Iberia, 2010 | 5.0 | 1 | 1 | 0 | 0 |
| *Total* | | 93 | 81(28) | 4 | 8 |

**D\*** is number of nodes where the target earthquakes have not been known before the recognition.

Totally, 93 strong earthquakes have occurred in the regions listed in Table 1; specifically, 81 events (or 87%) fall in the seismogenic nodes **D** recognized in advance, and 28 out of these 81 earthquakes are located within the nodes **D**, where target events (earthquakes with $M \geq M_0$) were not yet observed at the moment of the recognition. Thus the verification against independent data, i.e. the earthquakes that occurred after the publication of the maps, supports the reliability and statistical significance of the results of the recognition of the earthquake prone areas.

## 2. The morphostructural zonation and pattern recognition of earthquake prone areas in the Po plain (Northern Italy)

The recognition of earthquake prone areas was already carried out for the Italian mountain ranges, namely in the Alps (Gorshkov et al., 2004), peninsular Italy and Sicily (Gorshkov et al., 2002), using the pattern recognition approach. In this study, the analysis is extended to the flat areas in the Po Plain, to allow for the systematic identification of the nodes prone to earthquakes with magnitude $M \geq 5.0$.

### 2.1. Morphostructural zoning of the Po plain

The preliminary $MZ_P$ map of the Po plain, shown in Fig. 1, has been compiled using topographical, tectonic and geological maps as well as satellite photos. Geophysical and geological data, available via the portal of the Italian Geological Service on ISPRA website (http://www.isprambiente.gov.it), as well as the DISS database of active faults, have been considered. Recent publications on geomorphology, tectonics, seismotectonics and geophysical studies of the Po plain have been taken into consideration as well (Castiglioni, 1999; Livio et al., 2009; Toscani et al., 2009; Burrato et al., 2003; Handy et al., 2010; Michetti et al., 2012). The outlined map shows a hierarchical blocks-and-lineaments structure of the region and the locations of the morphostructural nodes that are formed around intersections or junctions of lineaments of different strike. This map has been used for the recognition of the nodes capable of earthquakes with $M \geq 5.0$, as described in the following.

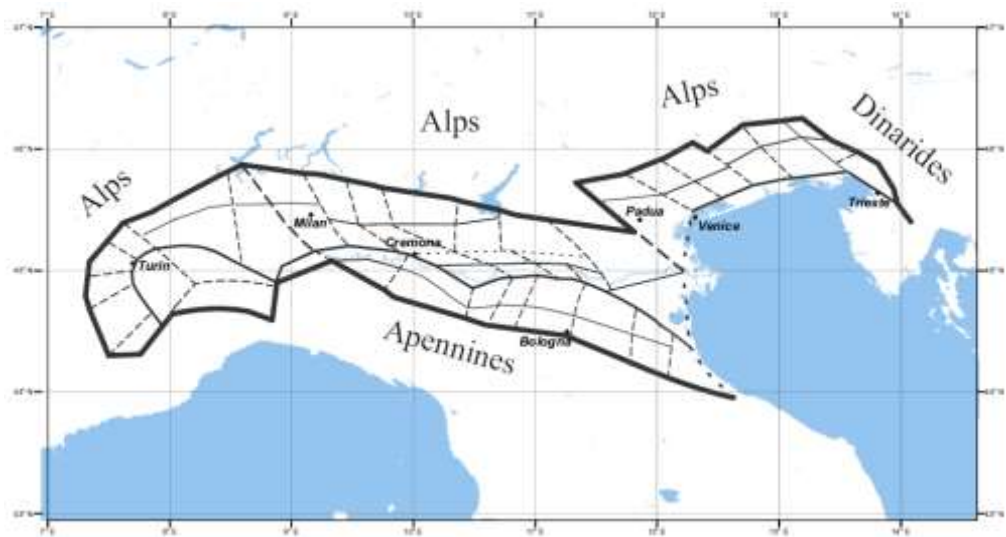

Fig.1. The morphostructural zoning map of the Po plain: lines outline morphostructural lineaments. Bold lines mark first rank lineaments, thick lines mark second rank lineaments, thin lines mark lineaments of third rank. Continuous lines correspond to longitudinal lineaments, while dashed lines indicate transverse lineaments.

The Po plain is characterized by a flat topography. Our recent studies of topographically flat basins in the Iberian Peninsula (Gorshkov et al., 2010) and the wider part of the Rhone Valley, between the Western Alps and Massif Central (Gorshkov and Gaudemer, 2012), have demonstrated the applicability of $MZ_P$ to the study of basins, with a relatively flat topography. The morphostructural analysis has been performed for the Po plain, following to the specific formal criteria applied to plain areas where, besides topography, special attention is paid to drainage pattern.

A preliminary correlation analysis between the epicenters of past of earthquakes with $M \geq 5.0$ and the intersections of $MZ_P$ lineaments has been performed. The information about earthquakes was selected from the UCI catalog (Peresan and Panza, 2002), that covers the time span from year 1000 up to 1 September 2012. We found that the considered events well correlate with intersections of the morphostructural lineaments, namely the morphostructural nodes, thus allowing us to apply pattern recognition algorithms for the identification of nodes prone to earthquakes with $M \geq 5.0$ in the Po Plain.

## 2.2. Identification of earthquake-prone areas (M≥5.0) in the Po Plain by the pattern recognition technique

Following a procedure similar to that developed and applied in the Rhone Valley, France (Gorshkov and Gaudemer, 2012), a preliminary identification of the areas prone to earthquakes with magnitude larger or equal to 5.0 (referred as *M5+* hereinafter) has been carried out for the Po plain.

The nodes have been defined based on the map shown in Fig. 1. In total, 102 intersections of lineaments are identified and each intersection is treated as a node. Formally the node is defined as circle of radius R=20 km surrounding each point of intersection of lineaments. Such node dimension is consistent with the dimension of earthquake sources with $M \geq 5.0$ (Wells and Coppersmith, 1994).

The scope of the recognition analysis is to classify all the nodes, delineated within the study region, into one of the following two classes:

1. class **D** containing the nodes where earthquakes *M5+* may occur (namely, the earthquake prone areas);
2. class **N** including the nodes where only smaller earthquakes may occur.

In this work, a two steps process, composed by a learning stage and a recognition stage is used to recognize the nodes prone to earthquakes with $M \geq 5.0$. At the stage of recognition of the seismogenic nodes no use is made of the information about past seismicity, since the recognition is based only on morphological, geological and geophysical parameters. At the learning stage, instead, past earthquakes are taken into account to identify the most informative parameters to be used for the recognition; in some cases, however, the learning stage is skipped, and the parameters identified for other regions are directly used (e.g. Gorshkov et al., 2004).

*Seismicity data*

To select the sample nodes for the learning stage, the information on the recorded events with *M*≥5.0 is taken into account, considering the UCI catalog (Peresan and Panza, 2002). This earthquake catalog covers the time span from 1000 year up to August 2012 and it is representative for events with *M*≥5.0 (M5+ earthquakes) from 1500 (Panza et al., 2010; Vorobieva and Panza, 1993). The location and intensity of the selected earthquakes has been cross-checked with the information provided by the NT4.1 (Camassi e Stucchi, 1997) and CPTI11 (Rovida et al., 2011) earthquake catalogs. Specifically, 71 earthquakes have been considered to select the nodes for the learning stage; 18 earthquakes are situated within the Po plain, while 53 are located along the boundaries between the Po plain and surrounding mountains. Their epicenters are mapped in Fig. 2 that shows how the considered earthquakes (well) correlate with intersections of morphostructural lineaments, which have been defined deliberately ignoring seismicity data. There is just one earthquake, occurred in 1894 (M=5.1), which cannot be associated with a specific node; its epicenter is located between nodes 41 and 42.

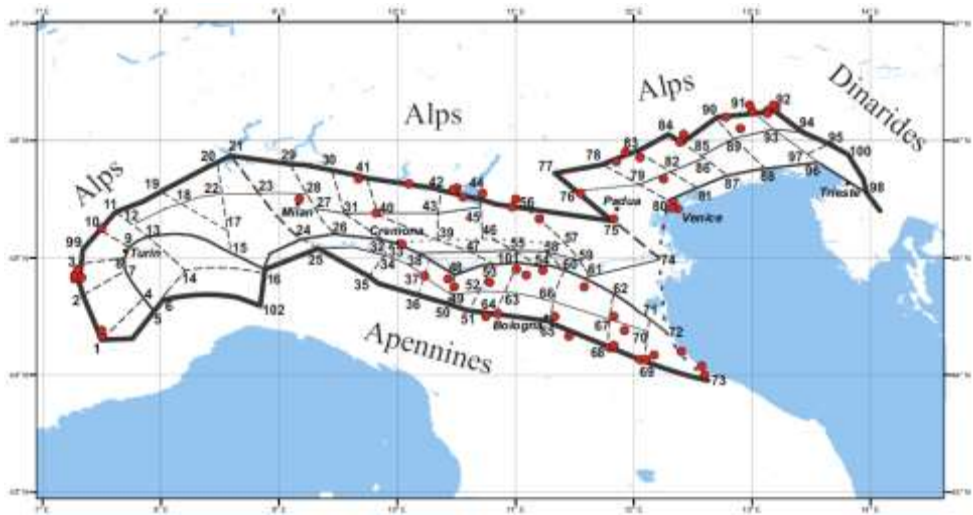

Fig. 2. Morphostructural zoning map of the Po plain and M5+ earthquakes occurred from 1093 to 2012. Lines depict lineaments of different rank (as in Fig. 1). Dots mark the epicenters of M5+ earthquakes from the UCI catalog.

*Definition of the parameters for the recognition process*

In order to apply the pattern recognition algorithms, each node is described by a set of parameters, the list of which is provided in Table 2. The values of the parameters have been measured for each node using topographic, geological and gravity maps, as well as the MZ map shown in Fig. 1.

**Table 2. Parameters used for recognition of seismogenic nodes in the Po plain and corresponding thresholds for their discretization**

| Parameters | Thresholds of discretization | |
| --- | --- | --- |
| | Po plain interiors | Po plain boundaries |
| *A)Topographic parameters* | | |
| Maximum topographic altitude, *m* (Hmax) | 30    20 | 901 1600 |
| Minimum topographic altitude, *m* (Hmin) | 5    30 | 70  180 |
| Relief energy, *m* (ΔH) (Hmax - Hmin) | 27    200 | 906 1560 |
| Distance between the points Hmax and Hmin, *km* (L) | 27    34 | 2   34 |
| Slope, (ΔH/L) | 2.1 | 32.2 61.8 |
| *B) Geological parameters* | | |
| The portion of the node area covered by soft (quaternary) sediments, % ,(Q) | 70    98 | 15  25 |
| *C) Parameters from the  morphostructural map* | | |
| The highest rank of lineament in a node, (HR) | 2 | 1 |
| Number of lineaments forming a node, (NL) | 2 | 2 |
| Distance to the nearest 1st rank lineament, *km,* (D1) | 30 | 0 |
| Distance to the nearest 2nd rank lineament, *km,* (D2) | 0  20 | 28 45 |
| Distance to the nearest node, *km,* (Dn) | 15    20 | 20 |
| *D) Morphological parameter (Mor)* | | |
| This parameter is equal to one of the following six values in accord with the morphology within each node: | 1.5 | 1.5 |
| 1 - mountain and plain (m/p) | | |
| 2 - piedmont and plain (pd/p) | | |
| 3 - plain only (p) | | |
| *E) Gravity parameters* | | |
| Maximum value of Bouguer anomaly, *mGal* ,(Bmax) | -65  -20 | -40 |
| Minimum value of Bouguer anomaly, *mGal,* (Bmin) | -110 -85 | -100  -60 |
| Difference between Bmax and Bmin, *mGal ,*(ΔB) | 30 50 | 20   40 |

*Recognition of nodes prone to earthquakes M5+*

Out of the 102 nodes identified within the study region, 60 are located in lowland environments within the Po plain and the remaining 42 sit on the first rank lineaments that divide the Po plain from the surrounding mountain chains. The considered parameters (Table 2), particularly the morphometric ones, differ significantly for the nodes located within the Po plain and for the nodes related with the first rank lineaments that delimit the plain. Therefore the recognition has been performed separately for these two groups of nodes, applying the CORA-3 recognition algorithm to each of them.

Accordingly, at the learning stage all the nodes have been a priori divided into two sets, so as to define the training sets for the CORA-3 algorithm. Specifically, to assemble the set **D₀**, the nodes situated most closely to the epicenters of past M5+ earthquakes are considered; the remaining nodes have been assigned to the set **N₀**.

*- Recognition of nodes located within the Po plain.* For the group of nodes located in the Po plain, four variants of classification of nodes into classes **D** and **N** have been performed, in order to identify the most stable one. The results of recognition are summarized in Table 3. Three variants of recognition (I-III) have been obtained using a different learning set **D₀**. In the variant I the set **D₀** is formed using the nodes located most closely to the epicenters of all earthquakes shown in Figure 1. In the variant II the nodes 28 and 80 (Fig. 2), hosting only ancient earthquakes whose location is poorly defined, are excluded from $\mathbf{D_0}$. Because the catalog UCI is representative for M5+ since 1500, in the variant III we exclude from $\mathbf{D_0}$ all the nodes hosting only events occurred before 1500 (specifically nodes 28, 33, 37, 80, 82, and 89). An additional classification of the nodes (IV) is obtained using the characteristic traits defined by CORA-3 for $\mathbf{D}$ and $\mathbf{N}$ nodes located in lowland areas of Iberia (Gorshkov et al., 2010). A last, additional experiment is carried out excluding from $\mathbf{D_0}$ the two nodes hosting the epicenters of the two earthquakes, which struck the Emilia Romagna region in May 2012 (i.e. nodes 54 and 101). Although the score of this last experiment is slightly less than those of variants I, II and III), both nodes are correctly recognized as $\mathbf{D}$ nodes. Therefore, if the recognition would have been performed any time before the event after the formalization of $MZ_P$ the node hosting the 20 May 2012 Emilia Romagna earthquake could have been properly recognized as prone to $M \geq 5.0$ earthquakes.

**Table 3. Results of recognition for nodes located in the Po plain. The total number of nodes, both D and N, is 60. The total number of past earthquakes M5+ occurred within the nodes is 18.**

| Variants of classification obtained with different learning sets $\mathbf{D_0}$ | **I** All events from UCI catalog are used to define $\mathbf{D_0}$ | **II** Nodes 28 (Milano) and 80 (Venice) are excluded from $\mathbf{D_0}$ | **III** Only nodes hosting events after 1500 are used to define $\mathbf{D_0}$ | **IV** **D-** and **N-**traits as defined for Iberia | **V** Nodes 54 and 101 (Emilia) are excluded from $\mathbf{D_0}$ |
|---|---|---|---|---|---|
| Total number of identified $\mathbf{D}$ nodes | **32** (53%) | **20** (33%) | **20** (33%) | **37** (62%) | **20** (33%) |
| Total number of target earthquakes within the identified $\mathbf{D}$ nodes | **18** (100%) | **16** (89%) | **13** (72%) | **8** (44%) | **12** (67%) |

The variants I and IV are discarded because too many nodes are assigned to class $\mathbf{D}$; moreover the variant IV also misses several nodes hosting target events. The variants II and III are practically equal, in spite of the difference in the learning sets $\mathbf{D_0}$. Although the variant III misses three nodes (namely 28, 80, 82), all hosting events occurred before 1500, it is considered the main variant, because, to define $\mathbf{D_0}$ only nodes hosting events $M5+$ that occurred after 1500 are used, when the epicentral locations in the catalogue UCI catalog are relatively more reliable. Nevertheless, the location of the nodes 28, 80 and 82 (Fig. 2), which experienced $M5+$ earthquakes before 1500 (when the reliability of the epicentral locations is severely conditioned by the naturally limited amount of available information), close to important cities (Venice and Milano) calls for a deeper investigation that will be performed in a forthcoming study.

The Po plain is generally characterized by moderate size earthquakes, with magnitudes not exceeding 5 for the majority of events. Nevertheless there are some important exceptions. In 2012 the south-east part of the Po plain, region Emilia-Romagna, was struck by the $M$=6.1 earthquake, which occurred at the node 54. The main classification of the nodes in the Po plain presented in Fig. 3 has been obtained using the node hosting the 2012 Emilia-Romagna earthquake for training the Cora 3 algorithm. To challenge the prediction capability of $MZ_P$ we have made an experiment excluding node 54 from the learning set $\mathbf{D_0}$. Even when it is excluded from $\mathbf{D_0}$, the node 54 is still recognized as $\mathbf{D}$.

*- Recognition of nodes related to first rank lineaments.*

Two variants of recognition, A and B have been obtained, using different learning sets $D_0$. In the variant A the set $D_0$ was formed using all the nodes situated most closely to the epicenters of past earthquakes M5+. In the variant B we excluded from $D_0$ three nodes 56, 75, 76 hosting events before 1500. The results of the recognition are presented in Table 4. In both variants, A and B, the same classification of the nodes into **D** and **N** classes is obtained. All the nodes hosting past earthquakes have been correctly recognized. Thus, each of the two variants can be considered as the main one.

The reliability of the recognition results has been evaluated by a set of control tests relevant to the determination of earthquake-prone areas (Gorshkov et al. 2003). We define as *main variants* the classification III presented in Table 3 and the classification B presented in Table 4. Four tests have been performed to analyze the stability of the nodes classification obtained for each group of nodes, e.g. checking the recognition of specific nodes when they are excluded from the training set or looking for possible equivalent traits. The results of the tests performed support the validity of the main variants.

Based on the described pattern recognition procedure the set of **D** nodes illustrated in Fig. 3 has been obtained, corresponding to the classification variant III for the Po plain and variant B for the nodes related to the first rank lineaments.

**Table 4. Results of recognition for nodes located along the boundaries (1st rank lineaments) between the Po plain and surrounding mountains. The total number of nodes, both D and N, is 42. The total number of past earthquakes M5+ occurred within the nodes is 53.**

| Classification variant | **A**<br>All events from the UCI catalog are used to select $D_0$ | **B**<br>Only nodes hosting events after **1500** are used to select $D_0$ |
|---|---|---|
| Total number of identified **D** nodes | **26**<br>(62%) | **26**<br>(62%) |
| Total number of target earthquakes within the identified **D** nodes | **52**<br>(98%) | **52**<br>(98%) |

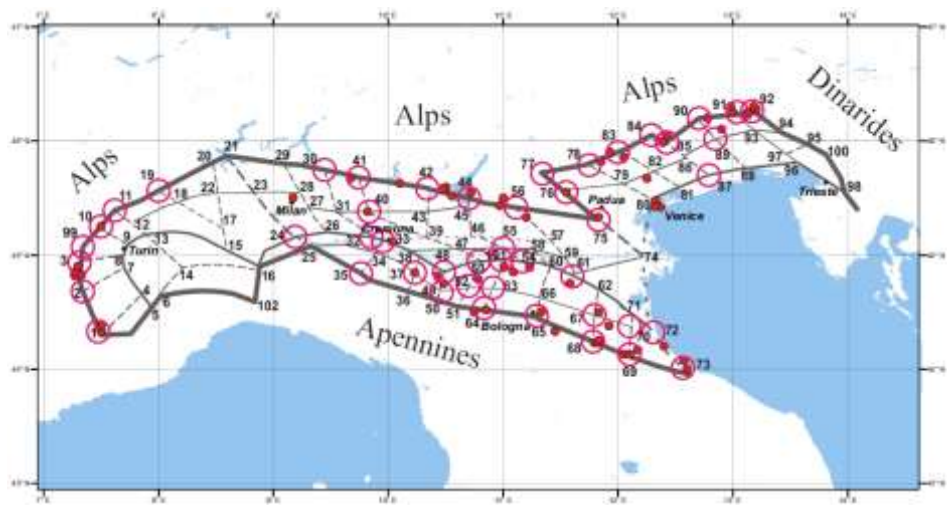

Fig. 3. Morphostructural zoning map of the Po plain and earthquakes M5+ from 1093 to 2012.
Lines depict lineaments of different rank. Dots mark epicenters *M5+* from the UCI0912 catalog. Red circles show recognized nodes prone to *M5+* earthquakes

## 3. Intermediate-term middle-range earthquake predictions based on precursory seismicity patterns

In addition to the recognition of areas prone to strong earthquakes (zero order or term-less prediction, Keilis-Borok and Soloviev, 2003) we show that the application of pattern-recognition techniques permits to carry out a quantitative analysis of the dynamics of seismicity, aimed at the space-time identification of the priority areas, which are characterized by an increased probability of strong earthquakes occurrence over the middle-range intermediate-term space-time scale.

Several possible scenarios of precursory seismic activity have been proposed; nevertheless, only a few formally defined algorithms allow for a systematic monitoring of seismicity, as well as for a widespread testing of their performances. Nowadays, one of the most promising approaches is represented by the intermediate-term middle-range earthquake predictions (i.e. with a characteristic alarm-time from a few months to a few years and a space uncertainty of hundreds of kilometers) based on the detection of formally defined variations in the background seismicity that precedes large earthquakes in a predefined area. In fact, the formal analysis of the seismic flow evidences that specific patterns in the events below some magnitude threshold, $M_0$, may prelude to an incumbent strong event, with magnitude above the same threshold $M_0$.

An essential step, when analyzing premonitory seismicity patterns, consists in the definition of the area where precursors have to be searched, the area of investigation, which is strictly interrelated with the size of the events to be predicted. For a natural reason, the size of the area should increase with the rupture size $L = L(M_0)$, where $M_0$ is the magnitude threshold. In particular, for the application of the methodology considered in this paper, it has been found empirically that the linear dimensions of the investigated area must be greater or equal to 5L-10L. This condition is especially relevant in view of the Multiscale Seismicity (MS) model (Molchan et al., 1997). In fact according to the MS model, the Gutenberg-Richter (GR) relation can be considered a law that describes adequately only the ensemble of earthquakes that are geometrically small with respect to the dimensions of the analysed region. In agreement with the MS model, when seismicity is analyzed over relatively small regions, the frequency-magnitude relation is linear (i.e. can be related with the Self-Organised Criticality paradigm, SOC) only up to a certain magnitude, while for the larger events it usually exhibits an upward bend. Specifically, within the areas delimited for prediction purposes, the number of earthquakes with $M>M_0$ (i.e. the target earthquakes, whose source size is comparable with the region size) usually exceeds the extrapolation of the GR law (i.e. does not follow SOC) and hence these events can be considered abnormally strong within the given region. At the same time the MS model guarantees the self-similarity condition for the small and moderate events ($M<M_0$) considered for the analysis of the premonitory patterns, and hence the log-linearity of the frequency-magnitude relation in the magnitude range of interest. In such a way, as discussed by Peresan et al. (2005), the algorithms make use of the information carried by the many small and moderate earthquakes, statistically obeing the GR law, to predict the strong earthquakes that do not satisfy the log-linearity of the GR law.

In this study two algorithms are considered, namely CN (Keilis-Borok and Rotwain, 1990) and M8 algorithms (Keilis-Borok and Kossobokov, 1990) that belong to a family of middle-range intermediate-term earthquake prediction algorithms based on a formalized, quantitative analysis of the seismic flow. The algorithms are based on a multiple set of premonitory patterns and have been designed following the general concepts of pattern recognition, which automatically imply strict definitions and reproducible prediction results. CN and M8 algorithms allow for a diagnosis of the Times of Increased Probability (TIPs) for the occurrence, inside a given region and time window, of earthquakes with magnitude greater than a fixed threshold $M_0$. Quantification of the seismicity patterns is obtained through a set of empirical functions of seismicity, each representing a reproducible precursor, whose definition has been guided by the theory of complex system and laboratory experiments on rocks fracturing. Specifically, the functions, which are evaluated on the sequence of the main shocks occurred within the analyzed region, account for increased space-time clustering of moderate size earthquakes, as well as for specific changes in seismic activity, including anomalous activation and quiescence.

The simple definition of alarm periods as "Times of Increased Probability with respect to normal conditions", which are not associated to a specific value of probability for the occurrence of a strong earthquake, is imposed by the fact that any attempt to quantify precisely the probability increase during TIPs would require several a priori assumptions (i.e. Poissonian recurrence, independence of TIPs, etc.), most of which would be poorly constrained by the available observations. Although many researchers still concentrate their efforts on assigning probability values, it is well known that making quantitative probabilistic claims, particularly for large and infrequent events, requires a long series of recurrences, which cannot be obtained at local scale from the existing catalogs of earthquakes. Resorting to subjective probability assessment, e.g. by expert elicitation process, may misled to the impression of detailed knowledge, which often turns out wrong (e.g. Kossobokov, 2005; 2008; 2009). Anyway decision-making does not necessitate a specific value of probability to be assigned, but rather requires an authoritative opinion on increased likelihood of incipient disaster (Guidelines for earthquake predictors, 1983). Thus, from a practical point of view, a hierarchical step-by-step approach to earthquake predictions (e.g. Kossobokov and Shebalin, 2003) appears preferable that ranges from term-less recognition of earthquake prone areas, to intermediate-term and, possibly, to short-term alerts. This multi-scale approach may benefit, at any stage, from independent complementary evidence from space geodesy, geochemical and other geophysical observations (Kanamori, 2003; Bormann, 2011, Panza et al. 2011).

The tests of predictions, performed on a global scale, already allowed for a statistical assessment of the predictive capability of M8 and CN algorithms, as confirmed by ICEF Report (Jordan et al., 2011). Specifically, for the M8 algorithm the on-going real-time prediction experiment, started in 1992 (Healy et al., 1992) for the great (M8.0+) and major (M7.5+) earthquakes worldwide (see http://mitp.ru/en/default.html), already demonstrated the high confidence level (above 99%) of the results (Kossobokov, 2012). For the algorithm CN a preliminary estimate of the significance of the prediction results, obtained for the period 1983-1998 in 22 regions of the world, provided a confidence level around 95% (Rotwain and Novikova, 1999). Such conclusions are supported by the statistically significant results obtained by the rigorous prospective testing of CN and M8S algorithms over the Italian territory (Peresan et al., 2011), ongoing for more than a decade, as described in this paper.

## 4. Prospective testing of CN and M8S predictions in Italy

Italy is the only region of moderate seismic activity where the two algorithms CN and M8S are currently applied simultaneously for the routine intermediate-term middle-range earthquake prediction (Peresan et al., 2005). Significant efforts have been made to minimize the intrinsic space uncertainty of predictions and the subjectivity of the definition of the areas where precursors should be identified (Peresan et al., 1999; Kossobokov et al., 2002). For the application of the algorithm CN, a regionalization strictly based on the seismotectonic zoning, and taking into account the main geodynamic features of the Italian area, is currently used (Peresan et al., 1999; Peresan e al., 2011), as shown in figure 4. For the application of the M8S algorithm (Kossobokov et al., 2002), that is a spatially stabilized variant of M8, the seismicity of the study region is analyzed within a dense set of overlapping circles, with radius increasing with increasing magnitude of the target events, that cover the monitored area. A hierarchy of predictions is delivered for different magnitude ranges $M_0+$, considering values of $M_0$ with an increment of 0.5 (i.e. $M_0+$ indicates the magnitude range: $M_0 \leq M \leq M_0+0.5$). In Italy M8S predictions are performed in the three different magnitude ranges defined by M6.5+, M6.0+ and M5.5+, with CIs of radius R=192 km, R=138 km and R=106 km respectively.

Several experiments have been dedicated to assess the robustness of the methodology against the unavoidable uncertainties in the data (Peresan et al., 2000; 2002). With these results acquired, a still ongoing experiment has been launched in July 2003, aimed at the rigorous real-time prospective testing of M8S and CN prediction for earthquakes with magnitude larger than a given threshold (namely 5.4 and 5.6 for CN algorithm, and 5.5 for M8S algorithm) in the Italian region and its surroundings. The goal of the experiment is to accumulate a collection of correct and wrong predictions (the latter include the false alarms and/or the failures to predict encountered in the test) permitting to verify and assess the predictive capability of the considered methodology. In this framework CN and M8S predictions have been routinely updated according to a predefined schedule, that is every two months CN predictions (i.e. on January, March, May, July, September, November) and every six months M8S predictions (i.e. on January and July). The rules for the real-time application of CN and M8S algorithms to the Italian territory are described in detail in Peresan et al. (2005).

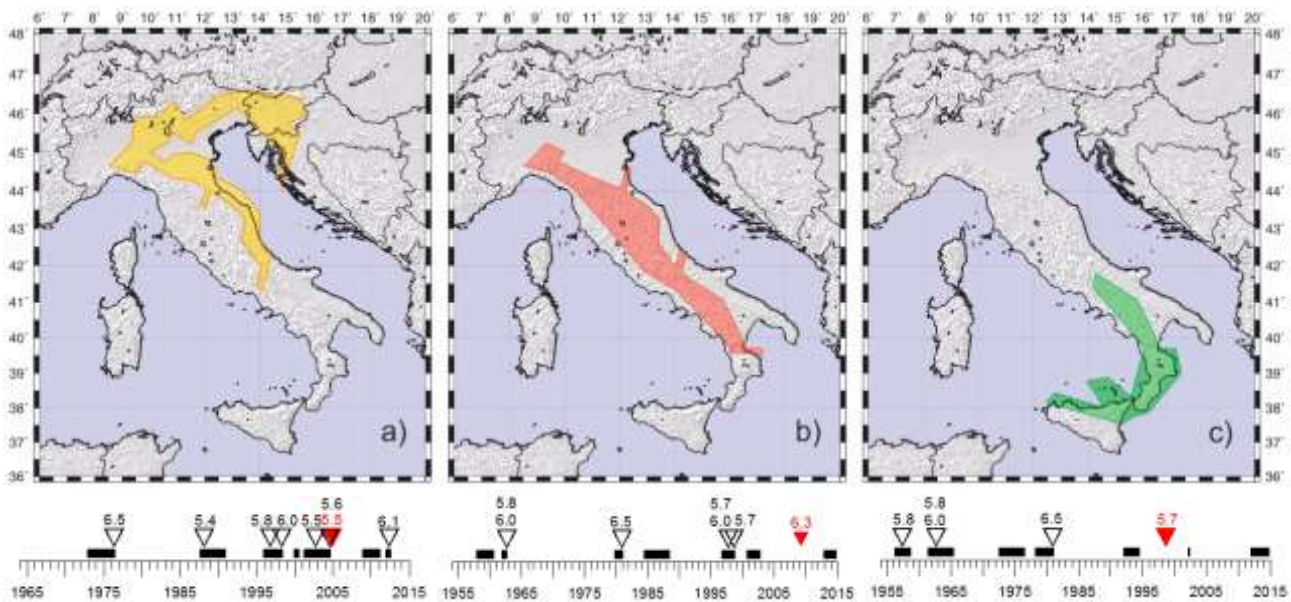

Fig. 4 - Regionalization used for the application of CN in Italy. In the diagrams of TIPs, the black boxes represent periods of alarm, while each triangle surmounted by a number indicates the occurrence of a strong event (M≥M$_0$), together with its magnitude. Triangles surmounted by two numbers indicate the occurrence of two target earthquakes within the same prediction time window (2 months). The magnitude thresholds are: a) Northern region M0=5.4; b) Central region M0=5.6; c) Southern region M$_0$=5.6. Full triangles indicate failures to predict.

The routinely updated predictions and a complete archive of predictions are made available on-line via the following website: http://www.geoscienze.units.it/esperimento-di-previsione-dei-terremoti-mt.html), thus allowing for rigorous validation and independent evaluation of the applied algorithms. The full list of target earthquakes for CN and M8S algorithms, both in retrospective and real-time application in Italy and adjacent territory, is provided in Table 5 and Table 6, respectively. The operating magnitude for M8S algorithm is the maximum magnitude, M$_{max}$, whereas for CN application M$_{prio}$ is considered, i.e., for each event, among the different magnitudes listed in the catalogue one is selected according to the priority criteria described in Peresan et al. (2005).

The results obtained so far, summarized in Tables 7 and 8, already permitted a preliminary assessment of the significance of the issued predictions by real-time monitoring (Peresan et al., 2011; 2012). All together 13 out of the 15 target earthquakes, occurred within the monitored territory since 1963, have been correctly preceded by an alarm (TIP, Time of Increased Probability) declared by CN algorithm. About 30% of the overall space–time volume (STV) is occupied by alarm and the confidence level of such predictions is above 99%. Similarly, the algorithm M8S correctly identified 14 of the 23 earthquakes with magnitude M5.5+ (i.e., between 5.5 and 6.0), occurred since 1972 within the monitored territory, with a STV of alarm of about 31%; the confidence level of M5.5+ predictions is above 98% (no estimation is yet possible for higher magnitude levels, due to the limited number of target events).

Particularly relevant appears the high confidence level of the real time prediction experiment. Specifically, since July 2003, 5 out of 7 target earthquakes have been predicted in real-time by CN with about 30% STV of alarms; M8S algorithm predicted 5 out of 9 earthquakes with a STV of alarms as low as 14% of the total monitored space-time volume; the probability of obtaining such results by random guess is below 3%. The space-time volume of alarms is computed accounting for the rate of occurrence of past earthquakes in the area under investigation (Relative Intensity, RI), following the procedure described by Kossobokov et al. (1999). The STV measure based on the seismicity distribution is more stringent than the measure based on simple spatial extent (in km) of alarm areas, and allows for the assessment of the significance of the issued predictions with respect to a time-independent and non-clustered seismicity model (i.e. RI), according to which future large earthquakes are considered more likely where higher seismic activity occurred in the past.

**Table 5. List of target earthquakes for CN retrospective (1954-1997) and real-time (1998-2014) application in Italy and adjacent territory. The earthquakes occurred during the period of forward monitoring, since 1998, are evidenced in bold**

| Date | Latitude, °N | Longitude,°E | Depth | $M_{prio}$ | CN | CN Region |
|---|---|---|---|---|---|---|
| 1957.05.20 | 38.70 | 14.10 | 60 | 5.8 | Yes | South |
| 1962.08.21 | 41.15 | 15.00 | 40 | 6.8 | Yes | Centre, South |
| 1962.08.21 | 41.15 | 15.00 | 40 | 6.0 | Yes | Centre, South |
| 1976.05.06 | 46.23 | 13.13 | 12 | 6.5 | Yes | North |
| 1980.11.23 | 40.85 | 15.28 | 18 | 6.5 | Yes | Centre, South |
| 1988.02.01 | 46.31 | 13.13 | 8 | 5.4 | Yes | North |
| 1996.10.15 | 44.79 | 10.78 | 10 | 5.8 | Yes | North |
| 1997.09.26 | 43.08 | 12.81 | 10 | 5.7 | Yes | Centre |
| 1997.09.26 | 43.08 | 12.81 | 10 | 6.0 | Yes | Centre |
| **1998.04.12** | **46.24** | **13.65** | **10** | **6.0** | **Yes** | **North** |
| **1998.09.09** | **40.03** | **15.98** | **10** | **5.7** | **Yes** | **Centre, South** |
| **2003.09.14** | **44.33** | **11.45** | **10** | **5.6** | **Yes** | **North** |
| **2004.07.12** | **46.30** | **13.64** | **7** | **5.7** | **Yes** | **North** |
| **2004.11.24** | **45.63** | **10.56** | **17** | **5.5** | **No** | **North** |
| **2009.04.06** | **42.33** | **13.33** | **8** | **6.3** | **No** | **Centre** |
| **2012.05.20** | **44.90** | **11.23** | **6** | **6.1** | **Yes** | **North** |

\* Only Central and Southern regions are considered up to 1964.

**Table 6. List of target earthquakes for M8S retrospective (1972-2001) and real-time (2002-2014) application in Italy and adjacent territory. Three magnitude ranges are considered: M6.5+, M6.0+, M5.5+. The earthquakes occurred during the period of forward monitoring are evidenced in bold**

| Date | Latitude, °N | Longitude,°E | Depth | M$_{max}$ | M8S | Commentary |
|---|---|---|---|---|---|---|
| **M6.5+** | | | | | | |
| 1976.05.06 | 46.23 | 13.13 | 12 | 6.5 | Yes | |
| 1980.11.23 | 40.85 | 15.28 | 18 | 6.7 | Yes | |
| **M6.0+** | | | | | | |
| 1997.09.26 | 43.08 | 12.81 | 10 | 6.4 | No | Predicted in M5.5+ |
| 1998.04.12 | 46.24 | 13.65 | 10 | 6.0 | No | Slovenia |
| **2009.04.06** | **42.33** | **13.33** | **8** | **6.3** | **No** | |
| **2012.05.20** | **44.90** | **11.23** | **6** | **6.1** | **No** | |
| **M5.5+** | | | | | | |
| 1975.01.16 | 38.20 | 15.78 | 21 | 5.5 | Yes | |
| 1978.04.15 | 38.27 | 15.10 | 18 | 5.8 | Yes | |
| 1979.09.19 | 42.72 | 12.95 | 6 | 5.8 | No | 20 km from alarm |
| 1980.05.28 | 38.46 | 14.34 | 19 | 5.5 | Yes | |
| 1986.11.25 | 44.12 | 16.34 | 30 | 5.5 | No | Croatia |
| 1990.05.05 | 40.78 | 15.77 | 10 | 5.6 | Yes | |
| 1990.11.27 | 43.85 | 16.63 | 24 | 5.6 | No | Croatia |
| 1991.02.26 | 40.19 | 13.82 | 401 | 5.5 | No | Deep event |
| 1994.01.05 | 39.08 | 15.15 | 272 | 5.8 | No | Deep event |
| 1996.10.15 | 44.79 | 10.78 | 10 | 5.8 | Yes | |
| 1997.09.26 | 43.05 | 12.88 | 10 | 5.9 | Yes | |
| 1998.05.18 | 39.25 | 15.11 | 279 | 5.6 | Yes | Deep event |
| 1998.09.09 | 40.03 | 15.98 | 10 | 5.9 | Yes | |
| 2001.07.17 | 46.73 | 11.20 | 10 | 5.5 | Yes | |
| **2002.09.06** | **38.38** | **13.70** | **5** | **5.9** | **No** | |
| **2002.10.31** | **41.79** | **14.87** | **10** | **5.9** | **No** | |
| **2003.03.29** | **43.11** | **15.46** | **10** | **5.5** | **Yes** | |
| **2003.09.14** | **44.33** | **11.45** | **10** | **5.6** | **Yes** | |
| **2004.02.23** | **47.27** | **6.27** | **17** | **5.5** | **Yes** | **Switzerland** |
| **2004.05.05** | **38.51** | **14.82** | **228** | **5.5** | **No** | **Deep event** |
| **2004.07.12** | **46.30** | **13.64** | **24** | **5.6** | **No** | **Slovenia** |
| **2004.11.24** | **45.63** | **10.57** | **24** | **5.5** | **Yes** | |
| **2006.10.26** | **38.67** | **15.40** | **216** | **5.8** | **Yes** | **Deep event** |
| **2009.04.06** | **42.33** | **13.33** | **6** | **6.3** | **No** | |

**Table 7. Space-time volume of alarm for CN application in Italy. N is the total number of target events, while n is the number of predicted earthquakes. During the period 1954-1963 only Central and Southern regions were analyzed.**

| Experiment | Space-time volume of alarm (%) | n/N | Confidence level (%) |
|---|---|---|---|
| Retrospective* (1954 – 1963) | 41 | 3/3 | 93 |
| Retrospective (1964 – 1997) | 27 | 5/5 | >99 |
| Forward (1998 – 2014) | 30 | 5/7 | 97 |
| All together (1954 – 2014) | 30 | 13/15 | >99 |

**Table 8. Space-time volume of alarm for M8S application in Italy, for the three magnitude ranges defined by M6.5+, M6.0+ and M5.5+. N is the total number of target events, while n is the number of predicted earthquakes. The confidence level of forward predictions for M5.5+ is above 99%.**

| Experiment | M6.5+ | | M6.0+ | | M5.5+ | |
|---|---|---|---|---|---|---|
| | Space-time volume of alarm (%) | n/N | Space-time volume of alarm (%) | n/N | Space-time volume of alarm (%) | n/N |
| Retrospective (1972-2001) | 35 | 2/2 | 39 | 1/2 | 38 | 9/14 |
| Forward (2002-2014) | 24 | 0/0 | 31 | 0/2 | 14 | 5/9 |
| All together (1972-2014) | 31 | 2/2 | 36 | 1/4 | 31 | 14/23 |

The uncertainties associated with intermediate-term middle-range earthquake predictions are intrinsically quite large; CN and M8S algorithms, however, already proved effective in predicting strong earthquakes, by rigorous prospective testing over the Italian territory. The increase in probability of strong earthquake occurrence associated with the alarmed areas (TIPs) can be grossly estimated based on the results obtained so far. The results from such analysis are summarized in Table 9. The yearly probability for a target earthquake occurrence within one of the CN monitored regions varies from 8% to 16%, depending on the region. When taking into account predictions, i.e. considering only the alerted time intervals (TIPs), such probability increases up to 24% and 48% respectively; accordingly, it is possible to quantify the probability gain associated with a TIP. Similarly, the probability for a target earthquake to occur, when no TIP is diagnosed, can be estimated around 2-3%. In all cases, the conditional probability, estimated accounting for the CN results, differs significantly from the average probability of target earthquakes occurrence within each of the monitored regions. This fact evidences the predictive capability of the algorithm. Moreover, in the diagram of time distribution of alarms (TIPs) reported in Fig. 4 it is possible to observe that the false alarm rate ranges from 30% for CN Northern region to about 50-60% in Southern region (Tab. 9); in several cases the false alarms are associated with the continuation of a TIP after the occurrence of a strong earthquake.

Based on similar reasoning, we have estimated the relative frequency of target earthquakes, which occurred within the entire space-time volume monitored by CN and M8S algorithms, and we have compared this estimate with that associated with the alerted volume (based on tables 7 and 8).

The frequency of target events turns out significantly higher within alarmed STV; namely, the frequency increases by about a factor 3 within the alarmed periods, when compared to the long term average rate. Accordingly, the probability gain associated with CN and M8S predictions for the Italian territory can be grossly estimated between 2 and 4, in good agreement with the independent estimates by ICEF (Jordan et al., 2011). Noticeably, these conclusions are based rigorously on the results from real-time prospective testing of the considered prediction algorithms, and thus reflect their effective predictive capability.

The operational effectiveness, along with the statistical validation, is a key element in the assessment of any earthquake prediction method; hence it should be properly defined and analyzed, avoiding approximate and incomplete representations, which often overlook the wide spectrum of possible mitigation actions (e.g. ICEF report, Jordan et al., 2011).

Although the predictive capability of the considered algorithms, with respect to a random Poissonian model, has been clearly recognized, it would be interesting to compare CN and M8S predictions against models based on earthquake clustering. However, so far no rigorously validated clustering model for the Italian territory is available to us to carry out such analysis, since the assessment for most of the clustering models is currently based on retrospective results and nominal probability gains. The prediction experiment by CN and M8S algorithms are, in fact, the only formally validated tools capable to anticipate the occurrence of strong earthquakes in the Italic region. CN and M8S prediction experiments started much earlier than the model testing within the Collaboratory for the Study of Earthquake Predictability – Testing Region Italy (i.e. August 2009; http://cseptesting.org/regions/italy) that has, so far, not released any document on the results achieved by prospective testing.

**Table 9. Analysis of the relative frequency of target earthquakes within the three regions monitored by CN algorithm (Fig. 4) in the period 1954-2014.**

|  |  | Number of events | Time (years) | Time % | Yearly probability % | Gain | False/Total alarms |
|---|---|---|---|---|---|---|---|
| **NORTH** | All time | 8 | 49,86 | 100 | 16 | 3,0 | 3/10 |
|  | **Alarm** | **7** | **14,66** | **29** | **48** |  |  |
|  | No Alarm | 1 | 35,20 | 71 | 3 |  |  |
| **CENTRE** | All time | 7 | 59,92 | 100 | 12 | 3,9 | 4/8 |
|  | **Alarm time** | **6** | **13,23** | **22** | **45** |  |  |
|  | No Alarm | 1 | 46,69 | 78 | 2 |  |  |
| **SOUTH** | All time | 5 | 59,92 | 100 | 8 | 2,9 | 5/9 |
|  | **Alarm time** | **4** | **16,73** | **28** | **24** |  |  |
|  | No Alarm | 1 | 43,19 | 72 | 2 |  |  |

### 4.1 Analysis of the stability of prediction results

In addition to the standard monitoring of precursory seismicity patterns, performed according to the rules defined in Peresan et al. (2005), a set of stability tests, with respect to the input earthquake catalogs, has been accomplished. A detailed description of the UCI earthquake catalog (Peresan et and Panza, 2002), used for the routine monitoring of seismicity by CN and M8S algorithms, is available at: http://www.geoscienze.units.it/esperimento-di-previsione-dei-terremoti-mt/the-italian-earthquakes-catalogue/124-description-of-the-catalogue-.html. Due to a remarkable incompleteness detected for the NEIC Preliminary Determinations of Epicenters since January 2009, the earthquake catalog UCI is currently updated using the NEIC data integrated by ISC bulletins information (starting on January 2012).

A preliminary comparative analysis of the different instrumental earthquake catalogs for the Italian territory, made available in the framework of the CSEP-TRI experiment (Collaboratory

Study on Earthquake Predictability, Testing Region Italy) has been carried out. The analysis evidenced the existence of relevant heterogeneities in the different instrumental data sets (Romashkova et al., 2009; Romashkova and Peresan, 2013), which should be taken into account when using them for the analysis of seismicity. The detected magnitude changes, in fact, may obscure possible precursory seismicity patterns or, in turn, may introduce spurious variations in reported seismicity, not related to any underlying physical process. The heterogeneous magnitude estimates provided over different time spans, therefore, prevent the use of the CSEP-TRI instrumental earthquake catalogs for CN and M8S algorithms testing in the Italian territory. The magnitude heterogeneity detected so far by Romashkova et al. (2009), has been recently corroborated by independent analysis by Gasperini et al. (2013). This observation poses serious concern on the results that could be obtained from CSEP-TRI testing based on such input data, as already evidenced by Peresan et al. (2012).

*Stability tests with respect to different input earthquake catalogs.*

Some experiments have been performed to assess the possibility of updating CN and M8S predictions using BSI data (Bollettino Sismico Italiano) since 2005. The BSI, in fact, is the authoritative catalog to be used for CSEP testing over the Italian territory (CSEP-TRI); remarkably such bulletins are available only starting from April 2005, whereas earlier data (i.e. bulletins from ISIDE website) are discontinuous and heterogeneous, as described by Romashkova and Peresan (2013).

Specifically, the variability of prediction results has been tested against the use of different input data, namely using the UCI catalog and the catalog obtained updating UCI by the Italian BSI bulletins since 16.04.2005 (referred as UCI+BSI in the following). The results of such experiments, provide essential information for the assessment of the possible operational use of BSI data for real-time prediction testing, as described hereinafter.

By the straightforward comparison of the annual number of earthquakes above different magnitude thresholds (Fig. 5), a clear magnitude discrepancy can be observed between the UCI and BSI data in the period of their overlapping, namely 2005-2012. The BSI magnitudes are systematically lower than those from UCI. The discrepancy seems to increase for larger earthquakes. The frequency-magnitude graphs (Fig.6) calculated for UCI for the periods before and after 16 April, 2005, and for BSI catalog since 16 April, 2005 confirm that BSI magnitudes are systematically lower than those reported in UCI catalog for the corresponding earthquakes.

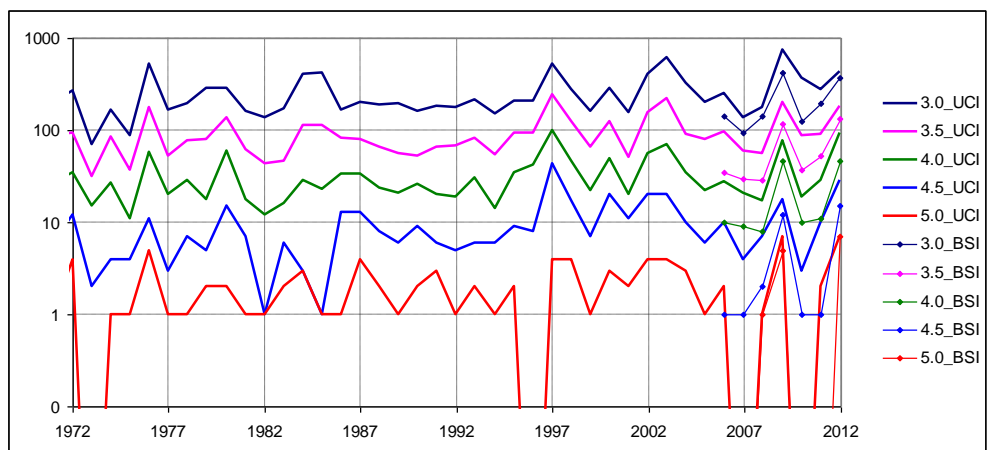

Fig. 5. Annual number of earthquakes above different magnitude thresholds versus time in the two datasets: UCI, 1972-2012, and BSI, 2005-2012. Both are selected for CSEP collection region, depth 0-30 km.

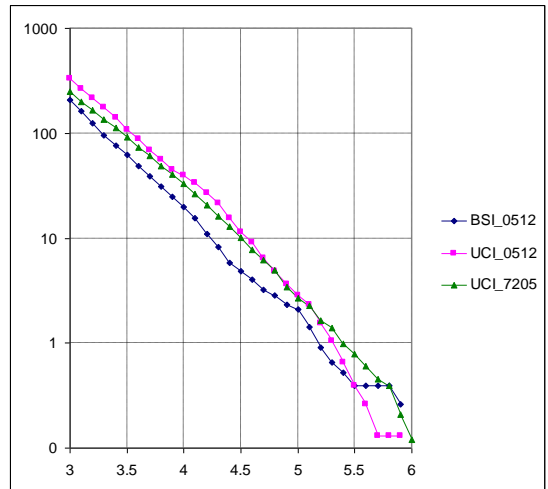

Fig. 6. Annual frequency-magnitude graphs for two catalogs in different time intervals: BSI, 16.04.2005-2012; UCI, 16.04.2005-2012, and UCI, 1972-15.04.2005.

Therefore it is natural to conclude that the straightforward updating of UCI by BSI starting from 16.04.2005 is not appropriate, in case the catalog is intended to be used for prediction purposes. Having this problem in mind, a test application of M8S algorithm has been performed using the UCI+BSI data, in order to check the variability of the results with varying input catalog. Since the area of CSEP region is smaller than that of standard M8S experiment for Italy and its surroundings, the M8S code has been modified to be applied to this restricted area. The results of the standard and modified M8S tests, where the data of UCI and UCI+BSI catalogs are used, show that the outputs are different, although some similarity in alarm areas exists. In general the space-time volume of alarm in M8S-CSEP test (i.e. considering CSEP region and UCI+BSI catalog) is smaller than that in the standard one, for all of the three magnitude ranges.

Similar results have been obtained for CN application using the composite UCI+BSI catalog. The analysis of the frequency-magnitude distributions within individual CN regions evidenced certain spatial heterogeneity; with data discrepancies particularly relevant within the CN Southern Region. As a result, when using UCI+BSI catalog an increase in the STV of alarms is observed for Central Region, while in Northern and Southern Regions the STV is reduced at the cost of additional failures to predict. Thus the use of BSI data causes a deterioration of prediction results, with respect to those obtained by the rigorous prospective testing, based on UCI catalog alone. Although the short time span of BSI data availability (about seven years) does not allow for a reliable assessment of the related predictions, the results of the preliminary data analysis and the test applications do not support the use of BSI data for the update of UCI on regular basis.

## 5. Neo-deterministic time-dependent seismic hazard scenarios for the Italian territory

A reliable and comprehensive characterization of expected seismic ground shaking, eventually including the related time information, is essential in order to develop effective mitigation strategies and increase earthquake preparedness. Moreover, any effective tool for seismic hazard assessment must demonstrate its capability in anticipating the ground shaking related with large earthquake occurrences, a result that can be attained only through rigorous verification and validation process against the real seismic activity.

The procedure for the neo-deterministic seismic zoning, NDSHA, is based on the calculation of realistic synthetic seismograms, as described in detail in Panza et al. (2012). Starting from the available information on the Earth structure, seismic sources, and the level of seismicity of the investigated area, ground motion is defined by full waveforms modelling. Hence the method does not make use of attenuation models (GMPE), which may be unable to account for the complexity of the medium and of the seismic sources (Paskaleva et al., 2007), and are often poorly constrained by the available observations (e.g. Burger et al., 1987; Panza and Suhadolc, 1989; Fäh and Panza, 1994; Bragato et al., 2011). NDSHA, in its standard form, defines the hazard as the maximum ground shaking at the site, computed considering a large set of scenario earthquakes, including the maximum credible earthquake, associated with the different sites. Seismic sources considered for ground motion modeling are defined based on the largest events reported in the earthquake catalogue, as well as incorporating the additional information about the possible location of strong earthquakes provided by the morphostructural analysis, active fault studies and other geophysical indicators (including GPS observations), thus filling in gaps in known seismicity (Panza et al., 2011).

Based on NDSHA approach, an operational integrated procedure for seismic hazard assessment has been developed that allows for the definition of time dependent scenarios of ground shaking, through the routine updating of earthquake predictions, performed by means of CN and M8S algorithms (Peresan et al., 2011). The integrated NDSHA procedure for seismic input definition, which is currently applied to the Italian territory, combines the different pattern recognition techniques, designed for the space-time identification of strong earthquakes, with algorithms for the realistic modeling of ground motion. Accordingly, a set of deterministic scenarios of ground motion at bedrock, proper for the time interval when a strong event is likely to occur within the alerted areas can be defined by means of full waveform modeling, both at regional and local scale.

In Italy and surrounding regions the areas prone to strong earthquakes (M≥6.0 and M≥6.5) have been systematically identified based on the morphostructural zonation and pattern recognition analysis (Gorshkov et al., 2002; 2004), according to the procedure described in Section 1. The identified seismogenic nodes are used, along with the seismogenic zones ZS9 (Meletti and Valensise, 2004), to characterize the earthquake sources used in the seismic ground motion modeling, as described by (Peresan et al., 2011). The earthquake epicenters reported in the catalogue are grouped into 0.2°x0.2° cells, and to each cell the maximum magnitude, recorded within it, is assigned. A smoothing procedure is then applied, to account for spatial uncertainty and for source dimensions. Only the sources located within the alarmed areas are considered to define the time-dependent scenarios.

Synthetic seismograms are then computed for sites placed at the nodes of a grid with step 0.2°x0.2° that covers the national territory, considering the average structural model associated to the regional polygon that includes the site. The seismograms used to be computed for an upper frequency content of 1 Hz, that is consistent with the level of detail of the regional structural models, and the point sources are scaled for their dimensions using the spectral scaling laws proposed by Gusev (1983), as reported in Aki (1987). From the set of complete synthetic seismograms, different maps of seismic hazard that describe the maximum ground shaking at the bedrock can be produced, including peak ground displacement (PGD), velocity (PGV) or design ground acceleration (DGA).

NDSHA has been recently extended to frequencies as high as 10 Hz; the preliminary results from this ongoing research (i.e. the regression relations between the strong motion parameters and the macroseismic intensities), basically confirm the results obtained with a 1 Hz cut-off frequency in the point-source approximation (Panza et al., 2012). On account of the available information about structural models and on the definition of damaging potential (Uang and Bertero, 1990; Decanini and Mollaioli, 1998) we consider, at the present stage of knowledge, the computation of velocities (10 Hz cut off frequency) the most stable and representative parameter to synthetically represent hazard. The scenarios of peak ground velocity associated with the three CN regions (Fig. 4) are illustrated in Figure 7.

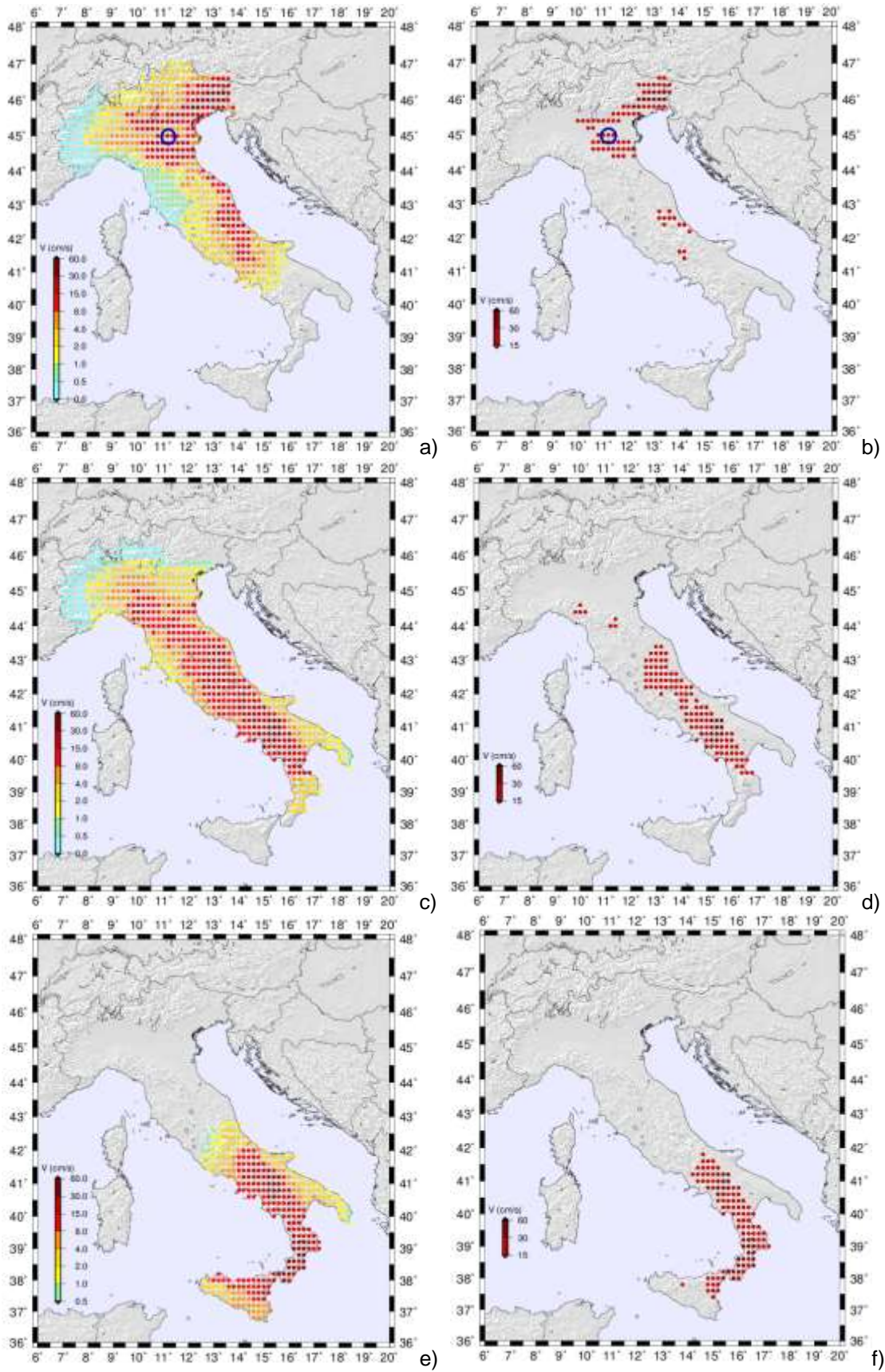

Fig. 7 – Time-dependent scenarios of ground shaking associated to an alarm in one of the monitored CN Regions (Fig. 4). On the left column acceleration maps are shown, computed using simultaneously all of the possible sources within the alarmed area and for frequencies up to 10 Hz. On the right, the same maps are provided, but for PGV>15 cm/s. The circle on maps a) and b) evidences the area within 30 km distance from the epicentre of the Emilia earthquake (M6.1, 2012). According to Tab. 9, when the alarm is declared, the probability associated with these scenarios ranges from 48% (Northern region,) to 24% (Southern region); when the alarm is off the corresponding probabilities are around 2-3%. Without accounting for predictions, the time-independent probabilities range from 8% to 16%, depending on the monitored region.

## 5.1 Prospective testing of time-dependent seismic hazard scenarios

CN and M8S predictions, as well as the related time-dependent ground motion scenarios associated with the alarmed areas, are routinely updated every two months since 2006. The procedure for the definition of the related ground shaking scenarios is illustrated in Peresan et al. (2011).

A strong earthquake (Mw=6.1) hit the Emilia region, Northern Italy, on 20th May 2012. The epicentre was localized inside the Northern Region (Fig. 4), alerted by CN algorithm for an earthquake with magnitude M≥5.4, starting on 1 March 2012, whereas it occurred outside the areas alerted by M8S algorithm for the corresponding magnitude interval. Therefore the earthquake scores as a successful real-time prediction, for CN algorithm only. The time-dependent ground shaking scenario associated to CN Northern region (Fig. 7a) defined for the period 1 March 2012 – 1 May 2012, correctly predicted the ground shaking, as large as ~0.25g, recorded for this earthquake. Notably, the ground shaking for this earthquake systematically exceeded the values expected at the bedrock in the area according to current Italian seismic regulation (i.e. PGA<0.175g), which is based on a classical PSHA map (Gruppo di Lavoro, 2004).

Since the time NDSHA time-dependent scenarios are regularly computed, namely starting on 2006, this is the second large earthquake that struck the Italian territory, along with L'Aquila earthquake (M=6.3, 2009). In both cases the method correctly predicted the observed ground motion, although L'Aquila earthquake scores as a failure in the earthquake prediction experiment, because the epicenter was located about ten km outside the alarmed territory (Panza et al., 2009; Peresan et al., 2011).

The results acquired in the prospective application of the time-dependent NDSHA approach allow for a rigorous validation of the integrated methodology and, at the same time provide information that can be useful to assign unbiased priorities for timely mitigation actions. As an example, for sites where ground shaking values greater than 0.2 g are estimated at bedrock (Fig. 7), further investigations can be performed taking into account the local soil conditions, to assess the performances of relevant structures, such as historical and strategic buildings (e.g. Vaccari et al., 2009) in addition to natural low key actions as described in Kantorovich and Keilis-Borok (1991).

The results presented in this section show that NDSHA permits to account for the temporal information about earthquakes occurrence provided by formally defined and tested tools for earthquake prediction. In addition, NDSHA can adequately address earthquake recurrence because it can naturally separate ground shaking from related recurrence (e.g. Panza et al., 2014), contrary to PSHA, which strictly depends on assumptions about the recurrence of large earthquakes that have large uncertainties and often turn out to be incorrect. Although temporal information may play a role in preparedness actions, a dilemma arises when accounting for it in the formulation of building codes. In PSHA, for example, the predicted ground shaking depends on the relative lengths of the time window used (i.e., on the probability threshold selected) and the recurrence time between large earthquakes. By forecasting the expected value of ground shaking over a time interval, PSHA underpredicts the actual shaking if earthquakes with longer recurrence times occur; such underestimates propagate non-linearly into unbearable errors in the expected human and economic losses (Wyss et al., 2012).

A comparative analysis among past seismicity, bedrock maps by NDSHA and by PSHA (reference for current seismic regulation) has been carried out for the Italian territory. Remarkably, the PSHA expected ground shaking estimated with 10% probability of being exceeded in 50 years (corresponding a return period of 475 years) appears underestimated (by about a factor 2) with respect to NDSHA estimates, particularly for the largest values of PGA (Zuccolo et al., 2011). At the same time, the probabilistic maps have a higher tendency to overestimate the hazard, with respect to available observations, particularly in low-seismicity areas. Therefore, the predictive capability of the PSHA maps turns out quite unsatisfactory, as shown by (Nekrasova et al., 2014).

## 6. Discussion and conclusions

We have shown, by rigorous prospective testing, that pattern recognition techniques proved effective in characterizing the space and time properties of impending strong earthquakes. The uncertainties associated with intermediate-term middle-range earthquake predictions are intrinsically quite large; however, they can be reduced by the combined use of independent information, including morphostructural and ground shaking information. Pattern recognition procedures, in fact, can be used to focus the investigation of possible local scale precursors in the areas (with linear dimensions of few tens kilometers), where the probability of a strong earthquake is relatively high, as well as for deriving time-dependent scenarios of ground shaking by the NDSHA approach (Peresan et al., 2011).

The recognized earthquake prone areas provide the first-order systematic information that may significantly contribute to reliable seismic hazard assessment in the Italian territory, as shown by Zuccolo et al. (2011). The new indications about the seismogenic potential obtained from this study, although less accurate than detailed fault studies, have the advantage of being independent from past seismicity, since they rely on the systematic and quantitative analysis of the available geological and morphostructural data. Even if past earthquakes are considered in the selection of traits typical for areas prone to large earthquakes (i.e. in the "learning stage"), the seismogenic nodes recognition is based on a set of parameters, like elevation and thickness of sediments, which do not include seismicity. Thus, this analysis appears particularly useful in areas where historical information is scarce; special attention should be paid to dangerous seismogenic nodes that are not yet related with known active faults or past earthquakes.

The information about the possible location of strong earthquakes provided by the morphostructural analysis can be directly incorporated in the neo-deterministic procedure for seismic hazard assessment, thus filling in possible gaps in known seismicity (Panza et al., 2012). Moreover, the space information about earthquake prone areas can be fruitfully combined with the space-time information provided by the quantitative analysis of the seismic flow, so as to identify the priority areas (with linear dimensions of few tens kilometers), where the probability of a strong earthquake is relatively high (Peresan et al., 2011; Panza et al., 2011), for detailed local scale studies.

The described procedure for the definition of time-dependent seismic hazard scenarios, based on pattern-recognition techniques and ground motion modeling, has been implemented within the framework of the SISMA project (http://sisma.galileianplus.it/) of the Agenzia Spaziale Italiana, an integrated prototype system for real-time joint processing of seismic and geodetic data streams (Crippa et al., 2008; Panza et al., 2011). The SISMA prototype is a fully formalized and highly automated tool (including version control for software and products) for the systematic real-time monitoring of deformations and seismicity patterns, which was made available to the Civil Defence of the Friuli Venezia Giulia Region for independent testing.

The routine application of the time-dependent NDSHA approach provides information that can be useful in assigning unbiased priorities for timely mitigation actions and, at the same time, allows for a rigorous prospective validation of the proposed methodology. The provided examples of the existing operational practice in predicting seismic ground shaking are perfectly in line, or even anticipating, the guidelines and recommendations given in the Report of the International Commission on Earthquake Forecasting (Jordan et al., 2011).

**Acknowledgements**

We are grateful to L. Romashkova for her fundamental contribution to the stability analysis of the intermediate-term middle-range earthquake predictions and related data. We acknowledge Giancarlo Neri and Mircea Radulian for their suggestions, which significantly contributed improving the text of the manuscript. This research benefited from the financial support by the DPC-INGV Agreement 2012-2013 Seismological Project S3 – "Previsione a breve termine dei terremoti". A.Gorshkov is partially supported by RFBR, grant 13-05-91167.